# Surface Electromagnetic Waves at Gradual Interfaces Between Lossy Media

Igor I. Smolyaninov

*Abstract*—A low loss propagating electromagnetic wave is shown to exist at a gradual interface between two lossy conductive media. Such a surface wave may be guided by a seafloor-seawater interface and it may be used in radio communication and imaging underwater. Similar surface waves may also be guided by various tissue boundaries inside a human body. For example, such surface wave solutions may exist at planar interfaces between skull bones and grey matter inside a human head at 6 GHz.

## I. Introduction

Abrupt step-like planar interfaces between media having different electromagnetic properties are known to sometimes support surface electromagnetic waves (SEW). However, long-range low-loss propagation of such SEWs is known to occur only in some limited circumstances. The most well-known cases of such low-loss SEWs include surface plasmons (SP), which propagate along metal-dielectric interfaces [1], and Zenneck waves [2], which may propagate along interfaces between highly lossy conductive media and low-loss dielectrics. In both cases the wave vector $k$ of such SEWs along the interface may be determined as:

$$k = \frac{\omega}{c}\left(\frac{\varepsilon_1 \varepsilon_2}{\varepsilon_1 + \varepsilon_2}\right)^{1/2} \quad (1)$$

where $\varepsilon_1$ and $\varepsilon_2$ are complex dielectric permittivities of the adjacent media, resulting in Im($k$)<<Re($k$) for both SPs and Zenneck waves [1,2]. In this paper we will consider a more general situation in which dielectric permittivity (or conductivity) of a medium change gradually across an interface. It appears that new kinds of SEW solutions may exist in such situations. We will demonstrate that a relatively low loss propagating SEW may exist at an interface between two highly lossy conductive media. While our analysis is applicable to any frequency range, we are primarily motivated by recent experimental progress in underwater radio communication [3] and potential application of our analysis in bioelectromagnetics.

## II. Theoretical Consideration

Let us solve macroscopic Maxwell equations in a non-magnetic ($\mu$=1) medium, in which the dielectric permittivity

I. I. Smolyaninov is with the Saltenna LLC, 1751 Pinnacle Drive, Suite 600 McLean VA 22102-4903 USA (phone: 443-474-1676; e-mail: igor.smolyaninov@saltenna.com) and with the University of Maryland (e-mail: smoly@umd.edu ).

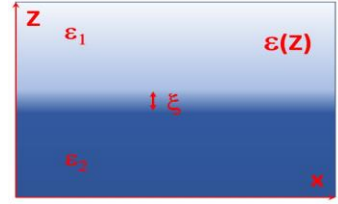

Fig. 1. Geometry of the problems. The dielectric permittivity of the medium $\varepsilon$ depends only on $z$ coordinate, which is illustrated by halftones. The gradual transition layer thickness between the two media equals $\xi$.

is continuous, and it depends on z coordinate only: $\varepsilon=\varepsilon(z)$, as shown in Fig.1. For such a geometry let us search for an electromagnetic wave propagating in the x direction, so that its field is proportional to $e^{i(kx-\omega t)}$. The macroscopic Maxwell equations may be written as

$$\vec{\nabla}\cdot\vec{D}=0, \ \vec{\nabla}\cdot\vec{B}=0, \ \vec{\nabla}\times\vec{E}=i\omega\vec{B}, \text{ and } \vec{\nabla}\times\vec{H}=-i\omega\vec{D} \quad (2)$$

which leads to a wave equation

$$\vec{\nabla}\times(\vec{\nabla}\times\vec{E}) = \frac{\omega^2 \varepsilon}{c^2}\vec{E} \quad (3)$$

After straightforward transformations this wave equation may be re-written as

$$-\nabla^2\vec{E} - \vec{\nabla}(E_z \frac{\partial \varepsilon}{\varepsilon \partial z}) = \frac{\varepsilon \omega^2}{c^2}\vec{E} \quad (4)$$

Depending on the polarization state (TE or TM) of the electromagnetic wave solution, Eq.(4) may be re-written in the form of the following effective Schrodinger equations:

$$-\frac{\partial^2 E_y}{\partial z^2} - \frac{\varepsilon(z)\omega^2}{c^2}E_y = -k^2 E_y \quad (5)$$

for the TE polarized wave (in which $E_z$=0), and

$$-\frac{\partial^2 E_z}{\partial z^2} - \frac{\partial E_z}{\partial z}\frac{\partial \ln \varepsilon}{\partial z} - \left(\frac{\varepsilon(z)\omega^2}{c^2} + \frac{\partial^2 \ln \varepsilon}{\partial z^2}\right)E_z = -k^2 E_z \quad (6)$$

for the TM polarized wave (in which $E_z \neq 0$). Eq.(6) may be further transformed by introducing the effective wave function $\psi$ as $E_z = \psi/\varepsilon^{1/2}$, which leads to

$$-\frac{\partial^2 \psi}{\partial z^2} + \left(-\frac{\varepsilon(z)\omega^2}{c^2} - \frac{1}{2}\frac{\partial^2 \varepsilon}{\varepsilon \partial z^2} + \frac{3}{4}\frac{(\partial \varepsilon/\partial z)^2}{\varepsilon}\right)\psi = -\frac{\partial^2 \psi}{\partial z^2} + V\psi = -k^2\psi \quad (7)$$

for the TM polarized wave. In both cases (Eq.(5) for the TE wave and Eq.(7) for the TM wave) $-k^2$ plays the role of effective energy in the respective Schrodinger equations. However, the effective potential energies in the TE and the TM cases are different. In the TE case the effective potential energy is $V(z) = -\frac{\varepsilon(z)\omega^2}{c^2}$, so that the only solutions of Eq.(5) having propagating character (Im($k$)<<Re($k$)) are those which behave like guided modes in a waveguide-like distribution of $\varepsilon=\varepsilon(z)$ (in which the dielectric permittivity is positive and almost pure real). Eq.(5) does not have SEW solutions.

Let us now study the TM-polarized solutions of Eq.(7). In agreement with the well-known SEW physics [1,2], surface plasmon and Zenneck SEW solutions do appear in this case due to the presence of derivative terms in the effective potential energy. For example, the $\frac{3}{4}\frac{(\partial\varepsilon/\partial z)^2}{\varepsilon^2}$ term dominates the potential energy expression in Eq.(7) when Re($\varepsilon$) gradually passes through zero at the metal-dielectric interface, leading to the appearance of surface plasmon SEW solutions. Indeed, this term becomes negative and strongly attractive whenever Re($\varepsilon(z)$) passes through zero:

$$V(z) \approx \frac{3}{4}\frac{(\partial\varepsilon/\partial z)^2}{\varepsilon^2} \approx -\frac{3}{4}\frac{(\partial \text{Re}(\varepsilon)/\partial z)^2}{\text{Im}(\varepsilon)^2} \quad (8)$$

However, careful examination of the effective potential energy expression in the TM case

$$V(z) = -\frac{4\pi^2\varepsilon(z)}{\lambda_0^2} - \frac{1}{2}\frac{\partial^2\varepsilon}{\varepsilon\partial z^2} + \frac{3}{4}\frac{(\partial\varepsilon/\partial z)^2}{\varepsilon^2} \quad (9)$$

(where $\lambda_0$ is the free space wavelength) reveals other situations which lead to novel propagating SEW solutions having Im($k$)<<Re($k$). Let us consider a gradual interface between two media, which have approximately the same loss angle $\delta$, and let us assume that the loss tangent remains approximately constant across the transition layer $\xi$ between these media (see Fig.1):

$$\varepsilon(z) = a(z)e^{i\delta}, \quad (10)$$

where $a(z)$ is the $z$-dependent magnitude of the dielectric permittivity. Under these assumptions the effective potential in the TM case (since $\omega = 2\pi c/\lambda_0$) equals

$$V(z) = -\frac{4\pi^2 a(z)e^{i\delta}}{\lambda_0^2} - \frac{1}{2}\frac{\partial^2 a}{a\partial z^2} + \frac{3}{4}\frac{(\partial a/\partial z)^2}{a^2} \quad (11)$$

If the spatial scale $\xi$ is much smaller than $\lambda_0$, the second and the third term will dominate in Eq.(11), and since these terms are real, Im($V$)<<Re($V$). The resulting potential well will be deep enough, so that the wave vector of the resulting surface wave solution will be very large ($k\sim 1/\xi>>2\pi/\lambda_0$), and this novel SEW will have propagating character (Im($k$)<<Re($k$)). Note that a 1D Schrodinger equation describing a one-dimensional potential well of arbitrary shape always have at least one eigenstate [4]. The SEW solution is tightly localized near the interface. Indeed, based on Eq.(7), far from the interface its electric field attenuates as $E_z\sim e^{-kz}$ away from the interface inside both media. On the other hand, in the opposite limit when $\xi$ is comparable or larger than $\lambda_0$, the newly found SEW solution disappears, since the first term will dominate in Eq.(11). This observation gives us the high frequency limit on SEW existence at a given fixed dielectric permittivity gradient.

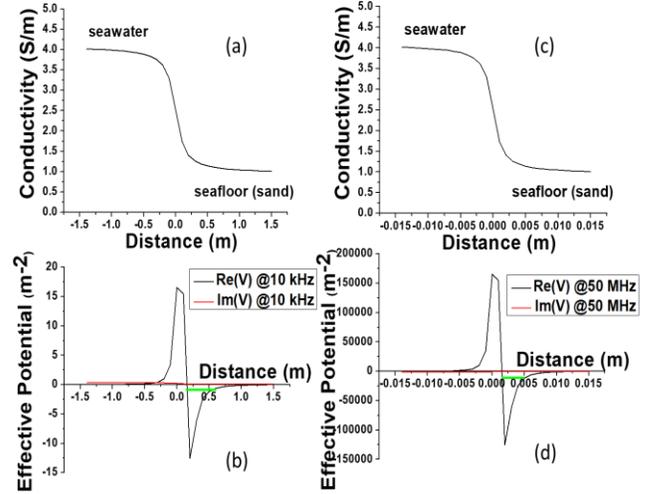

Fig. 2. (a) Plot of an assumed seabed $\sigma(z)$ transition layer in which conductivity changes from $\sigma_{water}$ to $\sigma_{sand}$ within about 0.5 m. (b) The corresponding effective potential energy (both real and imaginary parts) at the water-seabed interface defined by Eq.(13) is plotted for the 10 KHz band. The numerically obtained effective energy level is shown in green. (c) A more abrupt $\sigma(z)$ transition layer in which conductivity changes from $\sigma_{water}$ to $\sigma_{sand}$ within about 5 mm. (d) The corresponding effective potential energy at the interface is plotted for the 50 MHz band. The numerically obtained effective energy level is shown in green.

Another necessary condition of SEW existence established above is the approximate constancy of the loss angle $\delta$ across the interface. Otherwise, the Im($V$)<<Re($V$) condition will be violated, and the SEW solution given by Eq.(7) will not have propagating character. This condition gives us the limits on the frequency bands in which the newly found SEW solution may potentially exist for a given combination of the bounding media. The loss angle $\delta$ must remain approximately constant for both media within the limits of such band.

III. RESULTS OF NUMERICAL SIMULATIONS AND DISCUSSION

Let us confirm the analytical theoretical consideration above by detailed simulations in two situations of practical importance. First, let us analyze propagation of the newly found SEWs along the seawater-seabed interface and demonstrate that they may be used in long-distance underwater radio communication. The sandy seabed conductivity of $\sigma$=1 S/m has been measured in [5], while the average conductivity of seawater is typically assumed to be $\sigma$=4 S/m [3]. In the ELF-VHF radio frequency ranges the dielectric permittivities of seawater and sandy seabed are defined by their conductivities:

$$\varepsilon \approx i\varepsilon'' = \frac{i\sigma}{\varepsilon_0 \omega} \gg \varepsilon' \qquad (12)$$

where $\varepsilon_0$ is the dielectric permittivity of vacuum. Therefore, in these two media $\delta \sim \pi/2$ and it is safe to assume that $\delta$ remains approximately constant within the gradual transition layer between the seabed and the seawater. We will also assume that the large-scale roughness of the sandy seabed defines the width $\xi$ of the $\sigma(z)$ transition layer. Under these assumptions Eq.(11) may be re-written as

$$V = -\frac{i\sigma\omega}{\varepsilon_0 c^2} - \frac{1}{2}\frac{\partial^2 \sigma}{\sigma \partial z^2} + \frac{3}{4}\frac{(\partial \sigma/\partial z)^2}{\sigma^2} \qquad (13)$$

The effective potential energy defined by Eq.(13) is plotted in Fig.2 for two different cases of seabed surface roughness and two different operating frequencies (note that in both cases Im($V$)<<Re($V$)). The long-propagating-range eigenstate in both cases may be approximately determined using the virial theorem [4] as

$$k^2 \approx -\frac{\int_{-\infty}^{+\infty} \psi V(z)\psi^* dz}{2\int_{-\infty}^{+\infty} \psi \psi^* dz} \qquad (14)$$

(due to almost $\sim 1/z$ functional behavior of $V(z)$ near the potential barrier, see Fig.2(b,d)). However, to obtain better precision, in this work the effective Schrodinger equation was solved numerically using the Numerov method [6]. This method is the most commonly used method to numerically solve ordinary differential equations of second order in which the first-order term does not appear. The numerically obtained effective energy levels are shown schematically in green in Fig.2(b,d). The wavelength of the resulting surface wave solution is $\lambda = 2\pi/k$, and $L=$Im($k$)$^{-1}$ defines the SEW propagation distance. Based on the numerical solution of the resulting Schrodinger equation, it appears that a 10 kHz radio signal propagation distance in the case considered in Fig.2(a,b) reaches about $L\sim 50$ m, which considerably exceeds the classical skin depth of about 3 m at 10 kHz in seawater [3]. The surface wave propagation distance is about seven times larger than its wavelength computed numerically as $\lambda = 2\pi/k=7$ m, which clearly indicates "propagating" character ($\lambda \ll L$) of the surface wave. Using a good radio receiver, which are typically capable of operating down to at least $\sim$ -100 dBm signal levels, should allow communication distances of the order of 500 m at 10 kHz along the sandy seabed. If a sharper transition region from $\sigma_{water}$ to $\sigma_{sand}$ of the order of 5 mm is assumed, the effective potential (shown in Fig.2(d)) for the 50 MHz band gives rise to $\lambda = 2\pi/$Re($k$)=7 cm at this frequency, and the corresponding propagation length $L=1/$Im($k$) equals about 1 m. Under these conditions, using a good radio receiver capable of operating down to $\sim$ -100 dBm signal levels should allow communication distances of the order of 10 m along the sandy seabed.

Let us now consider SEW solutions which may exist at planar interfaces between different tissues inside a human body. As a general observation, commonly used FCC data on the dielectric parameters of various human body tissues [7] indicate that at 6 GHz the loss tangent variations between different tissues are not as pronounced as variations between the absolute magnitudes of $\varepsilon$. This observation is illustrated in Figs.3 and 4. In particular, the loss angle only varies by about 0.015 between the skull bone and the grey matter, while the absolute magnitude of $\varepsilon$ varies by approximately factor of 4.5 between these tissues. This observation indicates that under right conditions SEWs may be excited and guided by many various tissue boundaries inside a human body. As a proof of principle demonstration, and in order to avoid unnecessary

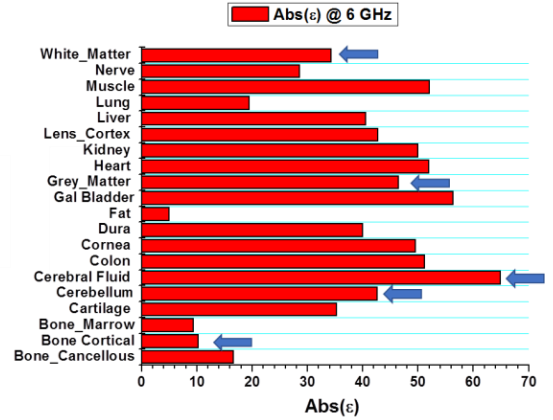

Fig. 3. Magnitude of the dielectric permittivity of various human tissues at 6 GHz (based on the data assembled in [7]). Various tissues present in the head are indicated by arrows.

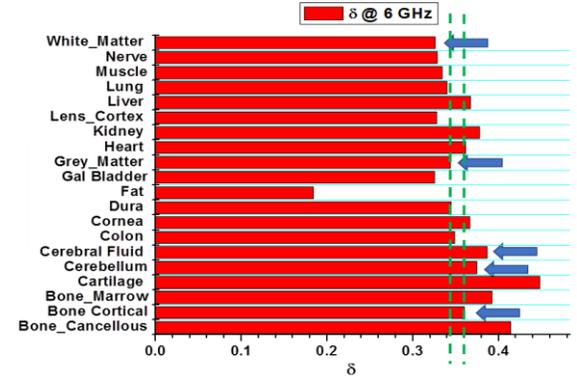

Fig. 4. Loss angle of various human tissues at 6 GHz (based on the data assembled in [7]). Various tissues present in the head are indicated by arrows. The green dashed lines indicate difference of the loss angle between the skull bones and the grey matter.

complications, let us consider an idealized geometry in which a planar skull bone is separated from the grey matter tissue by a 0.5 mm thick transition layer, in which the loss angle $\delta$ remains approximately constant at 6 GHz, as illustrated in Fig.5(a). The thickness of the transition region was chosen based on the lattice-like appearance of the bones of the cranial vault with a typical scale of the order of 500 μm - see for example Fig.3 from [8] (note also that the bone-tendon junctions also have similar transition layer thicknesses – see Fig.5 from [9]). A SEW solution having Im($k$)<<Re($k$) does appear in this case, and it is indicated by green line in Fig.5(b). Potential effects of SEW excitation and propagation along such an interface are illustrated in Fig.6, which depicts

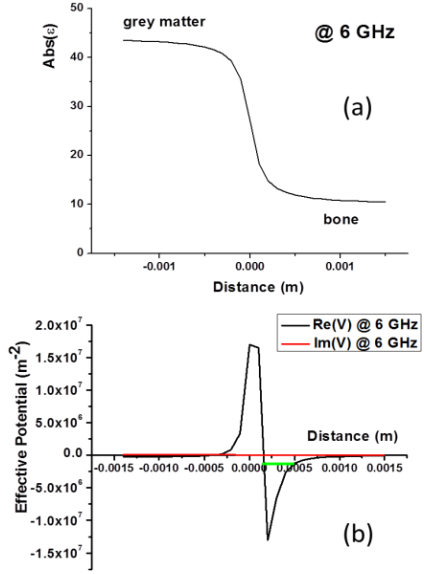

Fig. 5 (a) Plot of an assumed transition layer in which the magnitude of the dielectric permittivity changes from its value in the skull bone to its value in the grey matter within about 0.5 mm. (b) The corresponding effective potential energy (both real and imaginary parts) at the interface defined by Eq.(11) is plotted for the 6 GHz band. The numerically obtained effective energy level is shown in green.

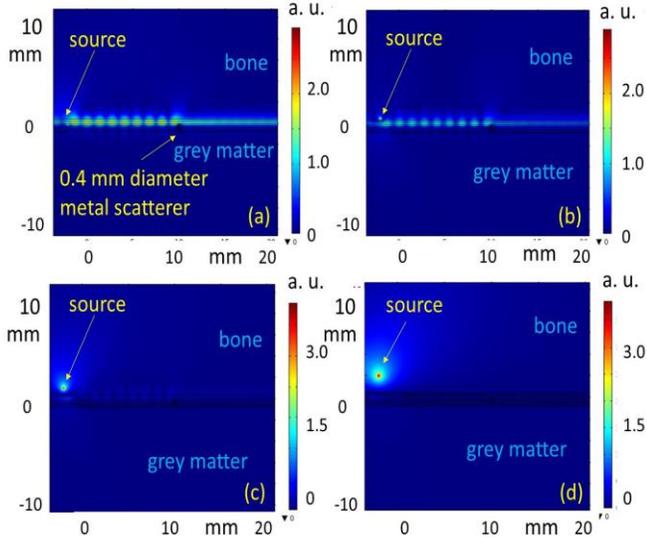

Fig. 6 Simulations of surface wave excitation and scattering in the transition layer between the skull bone and the grey matter at 6 GHz. The SEW electric field images are 25 mm x 25 mm. The distance between an excitation source and the transition layer increases progressively from (a) to (d). The SEW is excited by a 6 GHz point source only when the source is placed within about 1 mm from the interface. The RF field in the effective SEW waveguide is scattered by a 0.4 mm diameter metal particle placed inside the waveguide.

numerical simulations of surface wave excitation by a 6 GHz point source located at different distances from the interface. In order to illustrate the effects of SEW scattering, a 0.4 mm diameter metal particle has been placed inside the transition layer. These simulations indicate that the deeply subwavelength effective SEW waveguide, which is formed by the transition layer, only gets excited when the point source is placed in the immediate vicinity of the transition layer. The SEW is not excited when the point source is separated by more than ~ 1 mm from the transition layer. While this result is encouraging from the EM radiation safety prospective, simulations depicted in Fig. 7 look much less reassuring. In these simulations a SEW cavity is formed within the transition layer by placing two 0.4 mm diameter cylindrical metal scatterers inside the layer, which are separated by 2mm, which approximately equals one half of the SEW wavelength at 6 GHz (note that the SEW wavelength is much smaller than the free space wavelength at 6 GHz). As can be seen from Fig.7, under such circumstances the SEW cavity is excited rather efficiently even if the distance between the cavity and the point source reaches centimeter-scale distances.

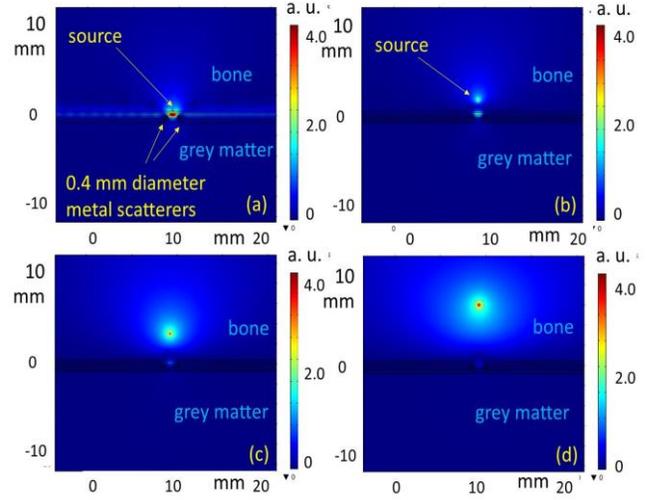

Fig. 7 Simulations of a SEW cavity excitation by a 6 GHz point source positioned at different distances from the interface. The distance between an excitation source and the interface increases progressively from (a) to (d). The SEW electric field images are 25 mm x 25 mm. The SEW cavity is formed by two 0.4 mm diameter cylindrical metal scatterers placed within the transition layer at half SEW wavelength (2mm) from each other. Under such circumstances the SEW cavity is excited rather efficiently even if the distance between the cavity and the point source reaches centimeter-scale distances.

IV. CONCLUSION

We have demonstrated that a low loss propagating surface electromagnetic wave may exist at a gradual interface between two lossy conductive media. Such a surface wave may be guided by a seafloor-seawater interface and it may be used in radio communication and imaging underwater. Similar surface waves may also be guided by various tissue boundaries inside a human body. For example, such surface wave solutions may exist at planar interfaces between skull bones and grey matter inside a human head at 6 GHz. A possibility of deeply sub-wavelength SEW cavities (or "hot spots") has been revealed in numerical simulations of SEW-related effects in human tissues. Since 6 GHz signals are widely used in the currently deployed 5G networks and novel wi-fi interfaces, it would be prudent to re-examine EM radiation safety issues associated with the potential excitation and scattering of the newly discovered surface electromagnetic waves inside a human body.

This work was supported in part by DARPA/AFRL Award FA8650-20-C-7027.